\documentclass[11pt]{article}
\usepackage{multicol}
\usepackage[utf8]{inputenc}
\usepackage{amssymb,amsmath}
\usepackage[hidelinks]{hyperref}
\usepackage{fixmath}
\usepackage[a4paper,hmargin=1.5cm]{geometry}

\def\R{\mathbb{R}}
\def\diffd{\mathrm{d}}
\let\phi\varphi

\title{A new approach to computing the asymptotics of the position of
Fisher-KPP fronts}

\author{Julien
Berestycki\footnote{\href{mailto:Julien.Berestycki@stats.ox.ac.uk}{\ttfamily Julien.Berestycki@stats.ox.ac.uk}, Department
of Statistics, University of Oxford, UK}, \'Eric
Brunet\footnote{\href{mailto:Eric.Brunet@lps.ens.fr}{\ttfamily Eric.Brunet@lps.ens.fr}, Laboratoire de
Physique
Statistique, \'Ecole Normale
Sup\'erieure, PSL Research University; Universit\'e Paris Diderot Sorbonne
Paris-Cit\'e; Sorbonne Universit\'e; CNRS.},
Bernard
Derrida\footnote{\href{mailto:Bernard.Derrida@lps.ens.fr}{\ttfamily Bernard.Derrida@lps.ens.fr}, Laboratoire de
Physique
Statistique, \'Ecole Normale
Sup\'erieure, Coll\`ege de France.}}

\begin{document}

\maketitle

\begin{abstract}

This paper presents a novel way of computing front positions in
Fisher-KPP equations. Our method is based on an exact relation between 
the Laplace transform of
the initial condition and  some integral functional of the front
position. Using singularity analysis, one can obtain
the asymptotics of the front position up to
the $\mathcal O(\log t/t )$ term.
Our approach is robust and
can be generalised to other
front equations.

\end{abstract}
\begin{multicols}{2}
\section{Introduction}

The goal of this letter is to present a novel way of computing the asymptotic
position of a front propagating into an unstable phase. The typical
equation we consider is the Fisher-KPP equation \cite{Fisher.1937,KPP.1937},
\begin{equation}
\partial_t h = \partial_x^2 h + h - h^2\qquad\text{(Fisher-KPP)},
\label{FKPP}
\end{equation}
but our method is general and can be adapted to a large class
of other reaction-diffusion equations.
An important feature of
\eqref{FKPP} is that the solution converges to a travelling wave:
for an initial condition $h_0\in[0,1]$ such that
$h_0(x)\to 1$ as $x\to-\infty$ and
$h_0(x)\to 0$ exponentially fast as $x\to\infty$, then 
\begin{equation}
h(\mu_t+ z,t) \to \omega(z),
\label{TW}
\end{equation}
where $\mu_t$ is the position of the front (we will choose  $\mu_t$ in
such a way that $h(\mu_t,t)=\frac12$ but other choices are possible) and $\omega(z)$ is the travelling
wave.

In a recent work \cite{BerestyckiBrunetDerrida.2017}, we have shown how to apply our
method to an equation looking like \eqref{FKPP}, but with the non-linear
term replaced by a free boundary condition. This allowed us to understand
in great detail how the large $t$ asymptotics of the position $\mu_t$
depends on the initial condition. In the present paper, we show that our
method is much more general and can be applied to a large variety of
non-linear equations such as \eqref{FKPP}.

The determination of the position $\mu_t$ of the Fisher-KPP equation has
attracted an uninterrupted attention \cite{AronsonWeinberger.1975, 
McKean.1975,Bramson.1978, Uchiyama.1978,
Bramson.1983,
BrunetDerrida.1997,
EbertvanSaarloos.2000,
MuellerMunier.2014,
BrunetDerrida.2015,
Henderson.2016,BBHR.2016,
NRR.2016,Cole.2017} since the equation was introduced in 1937
\cite{Fisher.1937,KPP.1937}. So far all the results were
obtained either by probabilistic methods, or 
by computing precisely how the shape $h(\mu_t+x,t)$ of the centred front
converges to the travelling wave, and then to determine $\mu_t$. Our method
is different. It 
consists in writing a relation between the initial condition $h_0$
and $\mu_t$. To do so, introduce
\begin{equation}
\phi(r,t)=\int_{\R}\diffd z \, h(\mu_t+z,t)^2 e^{rz},
\label{defphi}
\end{equation}
and
\begin{equation}
\Psi(r)=
\int_{\R}\diffd x\, h_0(x) e^{r x}.
\label{defpsi}
\end{equation}
Then,
our main relation (derived in Section~\ref{derive}) is
\begin{equation}
\Psi(r)
=\int_0^\infty \diffd t\, \phi(r,t)e^{r\mu_t-(r^2+1)t},
\label{main}
\end{equation}
for any $r$ small enough so that both sides converge.
Notice from \eqref{defphi} that $\phi(r,t)e^{r\mu_t}$ is  independent
of $\mu_t$. Therefore,
\eqref{main} holds in fact for an arbitrary choice of
$\mu_t$ and, by itself,  it is not sufficient to determine the position of the front.
However, when $\mu_t$ 
is the position of the front, we then have
\begin{equation}
\phi(r,t)\to \hat\phi(r)\text{\quad with }\hat\phi(r):=\int\diffd z\,
\omega(z)^2 e^{rz},
\label{isapos}
\end{equation}
for $r$ small enough, and we can evaluate the speed of that convergence.
This eventually allows to determine the first terms of the large $t$
asymptotics of~$\mu_t$.

\section{Velocity selection}
At this point, one can already understand from
\eqref{main} and \eqref{isapos} how the asymptotic velocity of the front
$v=\lim_{t\to\infty} \mu_t/t$ depends on the initial condition.

First assume that $\Psi(r)$ is  singular as $r\nearrow\gamma\le1$,
meaning (roughly speaking) that $h_0(x)$ decays as $e^{-\gamma x}$. Then,
obviously, the right-hand-side of \eqref{main} must also have be singular
as $r\nearrow\gamma$. This singularity must
come from the large~$t$ part of the integral, when $\phi(r,t)$ is nearly
equal to $\hat\phi(r)$ according to \eqref{isapos}. When $r<\gamma$, the
integral in \eqref{main} converges because
$r\mu_t-(r^2+1)t$ goes to $-\infty$ linearly in $t$.
As $r$ crosses $\gamma$, the integral becomes
singular because $r\mu_t-(r^2+1)t$  changes sign.
This means that $\mu_t\sim v t$ with $v$ such that $\gamma
v -(\gamma^2+1)=0$, which is the expected relation between the decay rate
$\gamma$ and the velocity $v$ when $\gamma<1$.

When $\Psi(r)$ has no singularity up to
$r=1$ (meaning that the initial condition decays ``fast''),
the velocity of the front cannot be larger than~2 (otherwise, there would
a singularity at some $\gamma<1$ solution to $\gamma v = \gamma^2+1$) so it must be equal to~2 as there are no
positive travelling waves of speed less than~2; this is also a well known
fact of the Fisher-KPP equation.

\section{Higher order corrections}

We have just seen that
the position of the singularity determines the velocity: $\mu_t\approx v t$; we are now going
to see that the nature of the
singularity gives the next order terms in $\mu_t$. Let us 
illustrate this method by focusing on the Ebert and van Saarloos term
\cite{EbertvanSaarloos.2000}.

Assume, for simplicity, that the initial condition decays fast enough for
$\Psi(r)$ as given by \eqref{defpsi} to be analytic at $r=1$.
Since Bramson's work
\cite{Bramson.1983}, it is known  that
\begin{equation}
\mu_t= 2t - \frac{3}2\log t +  a +  o(1),
\label{bram}
\end{equation}
and we want to estimate the $o(1)$. As a first attempt, let us look at what
happens as $r\nearrow1$ in \eqref{main} when $\phi(r,t)$ is
replaced by its limit $\hat\phi(r)$ and $\mu_t$ is given by $2t
- \frac{3}2\log t +  a$ for $t>t_0$, without any further corrective terms.
Then, with these substitutions, $\Psi(1-\epsilon)$ would be equal to
\begin{equation}
f(\epsilon)+\hat\phi(1-\epsilon)e^{(1-\epsilon)a}
\int_{t_0}^\infty\kern-.7em\diffd t\,
\frac{e^{-\epsilon^2t}}{t^{\frac32}}e^{\epsilon\frac32\log
t}
\label{fauxPsi}
\end{equation}
where $f(\epsilon)$, which corresponds to the integral from 0 to $t_0$, is
obviously analytic. On the other hand, the
integral above is an incomplete Gamma function, which one can expand in
powers of $\epsilon$ to obtain
$$A+B\epsilon+6\sqrt\pi\epsilon^2\log\epsilon+C\epsilon^2+
\mathcal O(\epsilon^3),$$
where $A$, $B$ and $C$ depend on $t_0$, but where the singular term in
$\epsilon^2\log\epsilon$ does not. (See also Section~\ref{smallexp}.)

Such a singular term cannot be actually present in the expansion of
$\Psi(1-\epsilon)$, because we know (from our choice of initial condition)
that $\Psi$ is analytic at~1.
As in the linear case \cite{BerestyckiBrunetDerrida.2017}, the only
possibility for the $\epsilon^2\log\epsilon$ term  to disappear,  
is that it is cancelled by another $\epsilon^2\log\epsilon$ term coming
from the $o(1)$   in \eqref{bram}. One
finds that this $o(1)$ term must be given, to leading order,
by the Ebert and van Saarloos term:
\begin{equation}
\mu_t= 2t - \frac{3}2\log t +  a -\frac{3\sqrt\pi}{\sqrt t}+\cdots
\label{EvS}
\end{equation}
Repeating the same procedure, one can notice that inserting 
$\mu_t= 2t - \frac{3}2\log t +  a -\frac{3\sqrt\pi}{\sqrt t}$ into
\eqref{main} leads to a $\epsilon^3\log\epsilon$ singular term in the
expansion. By a careful small $\epsilon$ expansion, one finds as
illustrated in Section~\ref{smallexp} that this term is cancelled by choosing
\begin{equation}
\mu_t= 2t - \frac{3}2\log t +  a -\frac{3\sqrt\pi}{\sqrt t}+
\frac98(5-6\log2)\frac{\log t}t+\cdots,
\label{next}
\end{equation}
and so on: each new term in the large $t$ expansion of
$\mu_t$ allows to remove a singularity in the small $\epsilon$ expansion
of $\Psi$, but introduces a new, weaker, singularity.

Remark that we started this analysis by requiring that $\Psi(r)$ is
analytic at $r=1$. In fact, this hypothesis is not needed: to obtain
\eqref{EvS}, the only requirement is that there is no
$\epsilon^2\log\epsilon$ term in the expansion of $\Psi(1-\epsilon)$:
$$\Psi(1-\epsilon)=A+B\epsilon+o(\epsilon^2\log\epsilon)$$
for some constants $A$ and $B$.
 (From
\eqref{defpsi}, this condition is satisfied if the initial condition decays
a bit faster than $x^{-3}e^{-x}$.)
Similarly, the $(\log t)/t$ term of
\eqref{next} requires that there is no $\epsilon^3\log\epsilon$ term in
$\Psi(r)$, that is that the initial condition decays a bit faster than
$x^{-4}e^{-x}$.

\medskip

At the beginning of the current section, we have replaced
$\phi(r,t)$ in \eqref{main} by its limit $\hat\phi(r)$ to obtain
\eqref{fauxPsi}. It is now time to
justify this simplification. The term we neglected until now is
\begin{equation}
\label{defDelta}
\Delta(r)=\int_0^\infty \diffd
t\,\Big[\phi(r,t)-\hat\phi(r)\Big]e^{r\mu_t-(r^2+1)t}.
\end{equation}
We claim that
\begin{equation}
\Delta(1-\epsilon)=\tilde A+\tilde B\epsilon+\tilde C\epsilon^2+\mathcal O(\epsilon^3),
\label{Delta}
\end{equation}
which means that the first singularity in the small $\epsilon>0$ expansion
of $\Delta(1-\epsilon)$ is smaller than $\epsilon^3$.
Then, the result~\eqref{next} still holds as it was obtained by suppressing
a singularity $\epsilon^3\log\epsilon$, bigger than $\epsilon^3$.

To justify \eqref{Delta}, we argue in Section~\ref{oneovert}
that, when $\mu_t$ is defined as the position where the front is $1/2$, one
has
\begin{equation}
\phi(r,t)=\hat\phi(r)+\mathcal O\Big(\frac1t\Big).
\label{1/t}
\end{equation}
Then, inserting \eqref{1/t} and Bramson's estimate \eqref{bram} for the
position $\mu_t$ of the front into \eqref{defDelta}, one obtains
$$\Delta(1-\epsilon)=\int_1^\infty\diffd t\, \frac{e^{-\epsilon^2 t+\frac32\epsilon\log t}}{t^{3/ 2}}\times
\mathcal O \Big(\frac1t\Big).$$
One checks directly that the integral on the right hand side 
satisfies~\eqref{Delta}.

\section{Derivation of (\ref{main})}\label{derive}

From its definition
\eqref{defphi}, it is obvious that $\phi(r,t)e^{r\mu_t}$ is independent of
the choice of $\mu_t$. Thus, it is sufficient to establish (\ref{main}) for
 $\mu_t=0$.
Define, for $r$ small enough,
\begin{equation}
g(r,t)=\int_\R\diffd x \,h(x,t)e^{rx}.
\end{equation}
(Of course $\Psi(r)=g(r,0)$ from \eqref{defpsi}.)
Then, from \eqref{FKPP} and \eqref{defphi} with $\mu_t=0$ one has
\begin{equation}
\partial_tg(r,t)=(1+r^2)g(r,t)-\phi(r,t)
\label{eqPsit}
\end{equation}
where we integrated by parts $\int\diffd x\,\partial_x^2h\, e^{rx}$.
One can solve
\eqref{eqPsit} to get
\begin{equation*}
g(r,t)=e^{(1+r^2)t}\Big[\Psi(r) - \int_0^t \diffd s\, \phi(r,s)
e^{-(1+r^2)s}\Big].
\end{equation*}
It only remains to show that
\begin{equation}
g(r,t)e^{-(1+r^2)t}\to 0 \quad\text{as $t\to\infty$}
\label{remains}
\end{equation}
to conclude.
The solution $h(x,t)$ to \eqref{FKPP} is smaller than $L(x,t)$, the
solution to the linearised equation $\partial_t
L(x,t)=\partial_x^2L(x,t)+L(x,t)$ with $L(x,0)=h_0(x)$. For any $\beta$
such that $\Psi(\beta)<\infty$,
\begin{align}
L(x,t) 
&= \int_{\R}\diffd y \,
h_0(y)e^t\,\frac{e^{-\frac{(x-y)^2}{4t}}}{\sqrt{4\pi t}}
\notag\\&=
\int_{\R}\diffd y \, h_0(y)e^{(1+\beta^2)t-\beta(x-y)}\,\frac{e^{-\frac{(x-y-2\beta t)^2}{4t}}}{\sqrt{4\pi t}}
\notag\\&\le
\frac{e^{(1+\beta^2)t}}{\sqrt{4\pi t}}e^{-\beta x}\Psi(\beta).
\label{bound}
\end{align}
Then, we write that $h(x,t)\le\min[1,L(x,t)]$. By using the bound
\eqref{bound}, one has
\begin{equation}
h(x,t)\le \begin{cases}1 & \text{if $x<d_{\beta,t}$},\\
\frac{e^{(1+\beta^2)t}}{\sqrt{4\pi t}}e^{-\beta x}\Psi(\beta)& \text{if
$x>d_{\beta,t}$},
\end{cases}
\end{equation}
where $d_{\beta,t}$ is the position where the second bound is also equal to
1.
Then, for $r<\beta$, one gets
$$g(r,t)\le \Big(\frac1r+\frac{1}{\beta-r}\Big)e^{r d_{\beta,t}}.$$
Using $e^{rd_{\beta,t}}=\Big(\frac{e^{(1+\beta^2)t}}{\sqrt{4\pi t}}
\Psi(\beta)\Big)^{r/\beta}$, this leads for $t>1$ to
\begin{equation}
g(r,t)\le C e^{r(\beta+\beta^{-1})t}
\end{equation}
for some constant $C$.
Choose furthermore $\beta\le1$. With $r<\beta$,
one checks that $r(\beta+\beta^{-1})<1+r^2$, and one
concludes that \eqref{remains} and \eqref{main} hold for all $r<1$ such
that $r<\sup\big[\beta;
\Psi(\beta)<\infty\big]$.

\section{Justification of (\ref{1/t})}
\label{oneovert}

With $\mu_t$ the position where the front is $1/2$, define $$\delta(x,t)=h(\mu_t+x,t)-\omega(x)$$
one obtains from~\eqref{FKPP} that
$$\begin{aligned}
\kern-1em\partial_t\delta &= \partial_x^2\delta + 2\partial_x\delta +(1-2\omega)\delta
-(2-\dot\mu_t)(\partial_x\delta+\omega')-\delta^2
\\&\approx  \partial_x^2\delta + 2\partial_x\delta +(1-2\omega)\delta
-(2-\dot\mu_t)\omega'
\end{aligned} $$
where one neglected two second order terms  (recall that $\delta\to0$ and
$2-\dot\mu_t\to0$). With $\mu_t\approx2t-\frac32\log t$, one expects
$(2-\dot\mu_t)\sim 3/(2t)$ for large times. This means that
$$\delta(x,t)\sim \frac3{2t}\eta(x)\qquad\text{as $t\to\infty$},$$
with $\eta(x)$ the unique solution to
$$\eta''
+2\eta'+(1-2\omega)\eta=\omega'\qquad\eta(0)=0\qquad\eta(-\infty)=0.$$
(The $\partial_t\delta=\mathcal O(t^{-2})$ term is also negligible
compared to $\delta$, so that $\delta$ satisfies a nonhomogeneous second order
linear equation. We eliminate other solutions by using
$\delta(0,t)=0$, and $\delta(\pm\infty,t)=0$.)

One checks that $\eta(x)\sim -A x^3e^{-x}$ for large $x$, so that
the difference
$$\begin{aligned}
\phi(r,t)-\hat\phi(r)&=\int\diffd
x\,e^{rx}\Big[h(\mu_t+x,t)^2-\omega(x)^2\Big]\\
&=\int\diffd x\,e^{rx}\delta(x,t)\Big[h(\mu_t+x,t)+\omega(x)\Big]
\end{aligned}$$
converges nicely for $r$ around 1 (and even up to $r=2-\epsilon$), so
that  one obtains \eqref{1/t}.

\section{A small \texorpdfstring{$\epsilon$}{epsilon} expansion}\label{smallexp}
To illustrate the methods used in the present paper to obtain the
asymptotic expansion of $\mu_t$, we give here (without going into the
details of the computation) the small $\epsilon$ expansion of
$$I=\int_0^{\infty}\diffd t\, e^{-\epsilon^2t+(1-\epsilon)(\mu_t-2t)},$$
where $\mu_t$ is an arbitrary function such that, as $t\to\infty$,
$$\mu_t=2t -\frac32\log t +a +\frac b{\sqrt t} +\frac{c\log
t+d}{t}+o(t^{-1}),$$
for arbitrary constants $a$, $b$, $c$, $d$.
One finds
$$\begin{aligned}
I=&A_0+A_1\epsilon+ 2e^a(b+3\sqrt\pi)\epsilon^2\log\epsilon 
+A_2\epsilon^2\\
&-3e^a(b+3\sqrt\pi)\epsilon^3\log^2\epsilon 
\\&+e^a\Big[\Big(15-\frac83c 
-18 \log2\Big)\sqrt\pi\\
&\qquad\qquad-(3\gamma_E+2a-1)(b+3\sqrt\pi)\Big]
\epsilon^3\log  \epsilon
\\&+A_3\epsilon^3+o(\epsilon^3),
\end{aligned}$$
with $\gamma_E$ the Euler constant. Notice that the singular terms only
depend on the asymptotic behaviour of $\mu_t$, while the regular terms
$A_0$, $A_1$, \ldots\ depend on the whole function $\mu_t$. For instance,
 $A_0=\int_0^\infty\diffd t\, e^{\mu_t-2t}$
and $A_1=-e^a2\sqrt\pi+\int_0^\infty\diffd t\, e^{\mu_t-2t}(2t-\mu_t)$.
The value of $A_0$ is obvious, the value of $A_1$ is maybe less obvious,
and $A_2$ and $A_3$ have complicated expressions.

To remove the singularities in the expansion of $I$, the only possible
choice is
$b=-3\sqrt\pi$ and $c=\frac98(5-6\log2)$.
\section{Conclusion}
In this letter, we have presented a new method to study the Fisher-KPP
equation. It relies on a single relation \eqref{main} between
the initial condition $h_0$ (through $\Psi$) and the position
$\mu_t$ of the front. A careful analysis of the singularities in \eqref{main}
leads to the large time asymptotics of the position of the front. 

In \cite{BerestyckiBrunetDerrida.2017,BrunetDerrida.2015}, we already used
a similar method to
study, respectively, a linear front equation with a free boundary or
on the lattice. The $(\log t)/t$ term was first identified, for the lattice
case in \cite{BerestyckiBrunet.2016}. The main progress of the present work is to show that this
method is not limited to linear fronts, but works also in the non-linear
case. Our main relation in
\cite{BerestyckiBrunetDerrida.2017} was
simpler than \eqref{main} because the term $\phi(r,t)$ was absent. However,
we argue in Section~\ref{oneovert} that $\phi(r,t)$ converges fast enough
as $t\to\infty$ for the large time analysis in
\cite{BerestyckiBrunetDerrida.2017} to apply
equally in the present setting, for the Fisher-KPP equation.

The method presented here is robust, and can be adapted to
a wide variety of front equations. If one writes an equation such as
$$\partial_t h =\partial_x^2 h + h -F(h),$$
with the $h^2$ term replaced by an arbitrary non-linearity, very little is
needed to make sure that \eqref{main} still holds (after changing the
definition of $\phi$). In fact, $F(h)$ could
even be a functional of $h$ rather than a function, and \eqref{main} holds
for instance for the non-local Fisher-KPP \cite{BerestyckiNadinPerthameRyzhik.2009}
$$\partial_t h =\partial_x^2 h + h -h\rho* h,$$
where $\rho>0$ is some well-behaved kernel with $\int\rho=1$. 
One could also work with equations discrete in space and/or time
\cite{BrunetDerrida.1997,BrunetDerrida.2015}.

\medskip

J.B. was partially supported by ANR grants ANR-14-CE25-0014 (ANR GRAAL)
and ANR-14-CE25-0013 (ANR NONLOCAL)

\end{multicols}

\begin{thebibliography}{10}
\providecommand{\url}[1]{{#1}}
\providecommand{\urlprefix}{URL }
\expandafter\ifx\csname urlstyle\endcsname\relax
  \providecommand{\doi}[1]{DOI \discretionary{}{}{}#1}\else
  \providecommand{\doi}{DOI \discretionary{}{}{}\begingroup
  \urlstyle{rm}\Url}\fi

\bibitem{Fisher.1937}
R.A. Fisher, The wave of advance of advantageous genes, Annals of Eugenics
  \textbf{7}, 355 (1937)

\bibitem{KPP.1937}
A.~Kolmogorov, I.~Petrovsky, N.~Piscounov, \'etude de l'équation de la
  diffusion avec croissance de la quantit\'e de mati\`ere et son application
  \`a un probl\`eme biologique, Bull. Univ. \'Etat Moscou, A \textbf{1}, 1
  (1937)

\bibitem{BerestyckiBrunetDerrida.2017}
J.~Berestycki, E.~Brunet, B.~Derrida, Exact solution and precise asymptotics of
  a {F}isher-{KPP} type front, Journal of Physics A: Mathematical and
  Theoretical \textbf{51}, 035204 (2017)

\bibitem{AronsonWeinberger.1975}
D.G. Aronson, H.F. Weinberger, Nonlinear diffusion in population genetics,
  combustion, and nerve pulse propagation, in \emph{Partial Differential
  Equations and Related Topics}, Lecture Notes in Mathematics volume 446
  (Springer-Verlag, 1975), pp. 5--49

\bibitem{McKean.1975}
H.P. McKean, Application of {B}rownian motion to the equation of
  {K}olmogorov-{P}etrovskii-{P}iskunov, Communications on Pure and Applied
  Mathematics \textbf{28}, 323 (1975)

\bibitem{Bramson.1978}
M.D. Bramson, Maximal displacement of branching {B}rownian motion,
  Communications on Pure and Applied Mathematics \textbf{31}, 531 (1978)

\bibitem{Uchiyama.1978}
K.~Uchiyama, The behavior of solutions of some non-linear diffusion equations
  for large time, Journal of Mathematics of Kyoto University \textbf{18}, 453
  (1978)

\bibitem{Bramson.1983}
M.D. Bramson, Convergence of solutions of the {K}olmogorov equation to
  travelling waves, Memoirs of the American Mathematical Society \textbf{44}
  (1983)

\bibitem{BrunetDerrida.1997}
E.~Brunet, B.~Derrida, Shift in the velocity of a front due to a cutoff,
  Physical Review E \textbf{56}, 2597 (1997)

\bibitem{EbertvanSaarloos.2000}
U.~Ebert, W.~van Saarloos, Front propagation into unstable states: universal
  algebraic convergence towards uniformly translating pulled fronts, Physica~D
  \textbf{146}, 1 (2000)

\bibitem{MuellerMunier.2014}
A.H. Mueller, S.~Munier, Phenomenological picture of fluctuations in branching
  random walks, Physical Review E \textbf{90}, 042143 (2014)

\bibitem{BrunetDerrida.2015}
E.~Brunet, B.~Derrida, An exactly solvable travelling wave equation in the
  {F}isher-{KPP} class, Journal of Statistical Physics \textbf{161}, 801 (2015)

\bibitem{Henderson.2016}
C.~Henderson, Population stabilization in branching brownian motion with
  absorption and drift, Communications in Mathematical Sciences \textbf{14},
  973 (2016)

\bibitem{BBHR.2016}
J.~Berestycki, E.~Brunet, S.C. Harris, M.~Roberts, Vanishing corrections for
  the position in a linear model of {FKPP} fronts, Communications in
  Mathematical Physics \textbf{349}, 857 (2016)

\bibitem{NRR.2016}
J.~Nolen, J.M. Roquejoffre, L.~Ryzhik.
\newblock Refined long time asymptotics for {F}isher-{KPP} fronts (2017).
\newblock \urlprefix\url{http://arxiv.org/abs/1607.08802}

\bibitem{Cole.2017}
C.~Graham.
\newblock Precise asymptotics for {F}isher-{KPP} fronts (2017).
\newblock \urlprefix\url{https://arxiv.org/abs/1712.02472}

\bibitem{BerestyckiBrunet.2016}
J.~Berestycki, E.~Brunet.
\newblock A note on the convergence of the {F}isher-{KPP} front centred around
  its $\alpha$-level (2016).
\newblock \urlprefix\url{http://arxiv.org/abs/1603.06005}

\bibitem{BerestyckiNadinPerthameRyzhik.2009}
H.~Berestycki, G.~Nadin, B.~Perthame, L.~Ryzhik, The non-local {F}isher-{KPP}
  equation: travelling waves and steady states, Nonlinearity \textbf{22}, 2813
  (2009)

\end{thebibliography}
\end{document}